\documentstyle[12pt]{article}
\begin{document}
\parskip=5pt plus 1pt minus 1pt

\begin{flushright}
{\bf LMU-06(a)/95}\\
{March 1995}
\end{flushright}

\vspace{0.2cm}

\begin{center}
{\Large\bf Effect of $K^{0}-\bar{K}^{0}$ Mixing on $CP$ Asymmetries \\
in Weak Decays of $D$ and $B$ Mesons}
\end{center}

\vspace{1cm}

\begin{center}
{\bf Zhi-zhong XING} \footnote{{\it Alexander von Humboldt} $~$ Research
Fellow} \\
{\sl Sektion Physik, Theoretische Physik, Universit$\ddot{a}$t
M$\ddot{u}$nchen,} \\
{\sl Theresienstrasse 37, D-80333 Munich, Germany} \footnote{E-mail:
Xing@hep.physik.uni-muenchen.de}
\end{center}

\vspace{1.6cm}

\begin{abstract}

Within the standard electroweak model we carry out an instructive analysis of
the effect of
$K^{0}-\bar{K}^{0}$ mixing on $CP$ asymmetries in some weak decay modes of
$D^{\pm}$ and
$\stackrel{(-)}{B}$$^{0}_{d}$ mesons. We point out that a clean signal of $CP$
violation
with magnitude of 2Re($\epsilon$) $\approx 3.3\times 10^{-3}$ can manifest in
the
semileptonic decays $D^{+}\rightarrow l^{+}\nu^{~}_{l}K_{S}$ ($K_{L}$)
vs $D^{-}\rightarrow l^{-}\bar{\nu}^{~}_{l}K_{S}$ ($K_{L}$). The $CP$
asymmetries are also
dominated by 2Re($\epsilon$) in the two-body nonleptonic decays
$D^{\pm}\rightarrow
(\pi^{\pm}, \rho^{\pm}, a^{\pm}_{1})+ (K_{S}, K_{L})$ and in the multi-body
processes
$D^{\pm}\rightarrow \pi^{\pm}+n_{0}\pi^{0}+n_{\pm}(\pi^{\pm}\pi^{\mp})+ (K_{S},
K_{L})$,
where $n_{0}$ and $n_{\pm}$ are integers. It is possible to observe such
$CP$-violating
signals with about $10^{7}$ $D^{\pm}$ events at $\tau$-charm factories or
hadron
machines. Finally we show that the $CP$ asymmetry induced by Re($\epsilon$) may
compete with
those from $B^{0}_{d}-\bar{B}^{0}_{d}$ mixing and final-state interactions
in the semi-inclusive and exclusive decays $B_{d}\rightarrow
X(c\bar{c})+(K_{S}, K_{L})$
on the $\Upsilon (4S)$ resonance.

\end{abstract}

\vspace{2cm}

\newpage

It has been known for 30 years that there exists $CP$ asymmetry in the
$K^{0}\Leftrightarrow \bar{K}^{0}$ transitions [1]. This effect, conveniently
parametrized
by $\epsilon$, is only of order $10^{-3}$ [2]. So far, no other evidence for
$CP$ violation has been unambiguously established. Many sophisticated
experimental
efforts, such as the programms of $B$ factories [3], $\tau$-charm factories [4]
and
higher-luminosity hadron machines [5], are being made to discover new signals
of $CP$
asymmetries beyond the neutral kaon system.

\vspace{0.2cm}

$CP$ violation in the $Q=2/3$ quark sector, particularly in the $D$-meson
system, is
complimentary to that in the $K$- and $B$-meson systems. The standard
electroweak model
predicts rather small $D^{0}-\bar{D}^{0}$ mixing ($\Delta
m^{~}_{D}/\Gamma_{D}\leq 10^{-4}$)
and $CP$-violating effects in weak decays of $D$ mesons (at the level of
$10^{-3}$
or smaller) [4, 6-8]. Since many exclusive $D$ decay modes are Cabibbo-favored,
there are
still possibilities to observe small $CP$ asymmetries with about $10^{7-8}$
events,
e.g., at the forthcoming $\tau$-charm factories. If new physics were to enhance
$D^{0}-\bar{D}^{0}$ mixing or penguin-mediated processes, then large $CP$
violation
could also manifest in the $D$-meson system.

\vspace{0.2cm}

In this note we shall examine the effect of $K^{0}-\bar{K}^{0}$ mixing on $CP$
asymmetries
in some semileptonic and nonleptonic $D^{\pm}$ decays with $K_{S}$ or $K_{L}$
in the final states.
Although this effect was noticed in a few previous works (see, e.g., refs.
[4,7,8]), an instructive
study of it has been lacking. We point out that $K^{0}-\bar{K}^{0}$ mixing
can give rise to a clean signal of $CP$ violation with magnitude of
2Re($\epsilon$)
in the semileptonic transitions $D^{+}\rightarrow l^{+}\nu^{~}_{l}K_{S}$
($K_{L}$) vs $D^{-}\rightarrow l^{-}\bar{\nu}^{~}_{l}K_{S}$ ($K_{L}$). The $CP$
asymmetries
are also dominated by 2Re($\epsilon$) in the two-body nonleptonic decays
$D^{\pm}\rightarrow (\pi^{\pm}, \rho^{\pm}, a^{\pm}_{1})+(K_{S}, K_{L})$ and in
the
multi-body processes $D^{\pm}\rightarrow \pi^{\pm} +n_{0}\pi^{0}
+n_{\pm}(\pi^{\pm}\pi^{\mp}) +(K_{S}, K_{L})$,
where $n_{0}$ and $n_{\pm}$ are integers. Measurements of such $CP$-violating
signals can be
carried out with about $10^{7}$ $D^{\pm}$ events at either $\tau$-charm
factories or hadron machines.
Finally we carry out an analysis of $CP$ violation in the semi-inclusive and
exclusive decays
$B_{d}\rightarrow X(c\bar{c})+(K_{S}, K_{L})$ on the $\Upsilon (4S)$ resonance.
It is shown that
the Re($\epsilon$)-induced $CP$ asymmetry may compete with those from
$B^{0}_{d}-\bar{B}^{0}_{d}$
mixing and final-state interactions, and thus cannot be neglected {\it a
priori}.

\vspace{0.2cm}

We begin with the generic decay modes $D^{\pm}\rightarrow X^{\pm}K_{S}$ or
$X^{\pm}K_{L}$,
where $X$ denotes a semileptonic or nonleptonic state. To be more specific, we
concentrate
upon the transitions occurring only through the tree-level quark diagrams (see,
e.g., fig. 1).
Some such transitions have large branching ratios ($\geq 1\%$) and are
measurable in the
near future. In the presence of $CP$ violation $|K_{S}\rangle$ and
$|K_{L}\rangle$ are given by
\footnote{For simplicity, we neglect the overall normalization factor
$1/\sqrt{2(1+|\epsilon|^{2})}$
for $|K_{S}\rangle$ and $|K_{L}\rangle$.}
$$
\begin{array}{lll}
|K_{S}\rangle & = & (1+\epsilon)|K^{0}\rangle + (1-\epsilon)|\bar{K}^{0}\rangle
 \; , \\
|K_{L}\rangle & = & (1+\epsilon)|K^{0}\rangle - (1-\epsilon)|\bar{K}^{0}\rangle
 \; ,
\end{array}
\eqno{(1)}
$$
where the complex parameter $\epsilon$ signifies deviation of the mass
eigenstates
from the $CP$ eigenstates. Due to mixing, $K_{S}$ in the decay
$D^{+}\rightarrow X^{+}K_{S}$ may come from either $K^{0}$ or $\bar{K}^{0}$, or
both of them.
The transition amplitudes of this type of decays are symbolically written as
$$
\begin{array}{lll}
A(D^{+}\rightarrow X^{+}K_{S}) & = & (1+\epsilon^{*})T_{a}e^{{\rm i}\delta_{a}}
+(1-\epsilon^{*})T_{b}e^{{\rm i}\delta_{b}} \; , \\
A(D^{-}\rightarrow X^{-}K_{S}) & = & (1-\epsilon^{*})T^{*}_{a}e^{{\rm
i}\delta_{a}}
+(1+\epsilon^{*})T^{*}_{b}e^{{\rm i}\delta_{b}} \; ,
\end{array}
\eqno{(2)}
$$
where $T_{a}$ and $T_{b}$ denote the hadronic amplitudes containing the
weak phases, and $\delta_{a}$ and $\delta_{b}$ are the corresponding strong
phases.
The signal of $CP$ violation in the decay rates is obtained, to lowest order of
$\epsilon$, as
$$
\begin{array}{lll}
{\cal A}(X^{\pm}K_{S}) & \equiv & \displaystyle\frac{\Gamma(D^{-}\rightarrow
X^{-}K_{S})
- \Gamma(D^{+}\rightarrow X^{+}K_{S})}{\Gamma(D^{-}\rightarrow X^{-}K_{S}) +
\Gamma(D^{+}\rightarrow X^{+}K_{S})} \; \\ \\
& = & \displaystyle\frac{\Delta_{K_{S}} - 2{\rm Im}\left (T_{a}T^{*}_{b}\right
)\cdot\sin (\delta_{b}-\delta_{a})}
{|T_{a}|^{2} + |T_{b}|^{2} + 2{\rm Re}\left (T_{a}T^{*}_{b}\right )\cdot\cos
(\delta_{b}-\delta_{a})} \; ,
\end{array}
\eqno{(3a)}
$$
where
$$
\Delta_{K_{S}} \; =\; 2{\rm Re}(\epsilon) \cdot\left
(|T_{b}|^{2}-|T_{a}|^{2}\right )
+ 4{\rm Im}(\epsilon) \cdot {\rm Re}\left (T_{a}T^{*}_{b}\right )\cdot \sin
(\delta_{b}-\delta_{a}) \;
\eqno{(3b)}
$$
stands for the effect of $K^{0}-\bar{K}^{0}$ mixing on the $CP$ asymmetry
${\cal A}(X^{\pm}K_{S})$.
In a similar way one can obtain the $CP$ asymmetry in $D^{\pm}\rightarrow
K^{\pm}X_{L}$:
$$
\begin{array}{lll}
{\cal A}(X^{\pm}K_{L}) & \equiv & \displaystyle\frac{\Gamma(D^{-}\rightarrow
X^{-}K_{L})
- \Gamma(D^{+}\rightarrow X^{+}K_{L})}{\Gamma(D^{-}\rightarrow X^{-}K_{L}) +
\Gamma(D^{+}\rightarrow X^{+}K_{L})} \; \\ \\
& = & \displaystyle\frac{\Delta_{K_{L}} + 2{\rm Im}\left (T_{a}T^{*}_{b}\right
)\cdot\sin (\delta_{b}-\delta_{a})}
{|T_{a}|^{2} + |T_{b}|^{2} - 2{\rm Re}\left (T_{a}T^{*}_{b}\right )\cdot\cos
(\delta_{b}-\delta_{a})} \;
\end{array}
\eqno{(4a)}
$$
with
$$
\Delta_{K_{L}} \; =\; 2{\rm Re}(\epsilon) \cdot\left
(|T_{b}|^{2}-|T_{a}|^{2}\right )
- 4{\rm Im}(\epsilon) \cdot {\rm Re}\left (T_{a}T^{*}_{b}\right )\cdot \sin
(\delta_{b}-\delta_{a}) \; .
\eqno{(4b)}
$$
Note that $\delta_{b}\neq \delta_{a}$ is a necessary condition for nonvanishing
direct $CP$ violation in
the decay amplitude. In some previous studies,
$\Delta_{K_{S}}=\Delta_{K_{L}}=0$ was assumed. Subsequently we shall
take a few decay modes for example to illustrate the significant role of
Re($\epsilon$) in the $CP$ asymmetries.

\vspace{0.2cm}

{\large\it Example 1}. $~$ We first consider the semileptonic decays
$D^{+}\rightarrow l^{+}\nu^{~}_{l}K_{S}$ ($K_{L}$)
vs $D^{-}\rightarrow l^{-}\bar{\nu}^{~}_{l}K_{S}$ ($K_{L}$). In this case,
$T_{a}=0$ and the $CP$ asymmetries turn out to be
$$
{\cal A}(l^{\pm}K_{S}) \; =\; {\cal A}(l^{\pm}K_{L}) \; =\; 2 {\rm
Re}(\epsilon) \; =\;
2|\epsilon|\cos\phi_{\epsilon} \; .
\eqno{(5)}
$$
Clearly these asymmetries are equal in magnitude to that in $K_{L}\rightarrow
l^{+}\nu^{~}_{l}\pi^{-}$
vs $K_{L}\rightarrow l^{-}\bar{\nu}^{~}_{l}\pi^{+}$. The current experimental
data [2] give
$|\epsilon|\approx 2.27\times 10^{-3}$ and $\phi_{\epsilon}\approx 43.6^{0}$,
unambiguously leading to
${\cal A}(l^{\pm}K_{S})={\cal A}(l^{\pm}K_{L})$ $\approx 3.3\times 10^{-3}$.

\vspace{0.2cm}

{}From the averaged branching ratios of $D^{+}\rightarrow
l^{+}\nu^{~}_{l}\bar{K}^{0}$ ($l=e$ or $\mu$) [2],
one estimates Br($D^{+}\rightarrow l^{+}\nu^{~}_{l}K_{S}$) $\approx$
Br($D^{+}\rightarrow
l^{+}\nu^{~}_{l}K_{L}$) $\approx 3.4\%$. Thus observation of ${\cal
A}(l^{\pm}K_{S})$
to 3 standard deviations needs about $2.4\times 10^{7}$ $D^{\pm}$ events, if we
assume perfect
detectors or $100\%$ tagging efficiencies. In practice, there are two
possibilities to
reduce half of the total $D^{\pm}$ events needed for measuring the $CP$
asymmetry.
One way is to sum over the final states $e^{+}\nu^{~}_{e}K_{S}$ and
$\mu^{+}\nu^{~}_{\mu}K_{S}$
as well as their charge-conjugated counterparts. The other is to sum over the
final states
$l^{+}\nu^{~}_{l}K_{S}$ and $l^{+}\nu^{~}_{l}K_{L}$ as well as their
charge-conjugated counterparts.
The result in eq. (5) implies that both ways should not induce dilution of the
$CP$-violating signal.
A combination of these two ways is in principle possible too, then only about
$6\times 10^{6}$ events of
$D^{\pm}$ mesons are required to observe ${\cal A}(l^{\pm}K_{S})$.

\vspace{0.2cm}

{\large\it Example 2}. $~$ Now we consider $CP$ violation in the two-body
nonleptonic decays
$D^{\pm}\rightarrow (\pi^{\pm}, \rho^{\pm}, a^{\pm}_{1})+(K_{S}, K_{L})$. These
transitions can occur
through four tree-level quark diagrams, as shown in fig. 1. One expects that
the
annihilation-type diagram fig. 1(d) is formfactor-suppressed and thus
negligible. Regardless of
final-state interactions, the relative size of $T_{a}$ and $T_{b}$ in each
decay can be roughly
estimated by using the effective weak Hamiltonian and factorization
approximation [9]. Respectively
for $X=\pi, \rho$ and $a_{1}$, we obtain
$$
\frac{T_{b}}{T_{a}} \; = \; \frac{V_{cs}V^{*}_{ud}}{V_{cd}V^{*}_{us}}
\times \left \{
\begin{array}{ll}
& \displaystyle
1+\frac{a_{1}}{a_{2}}\cdot\frac{f_{\pi}}{f_{K}}\cdot\frac{F^{BK}_{0}(m^{2}_{\pi})}
{F^{B\pi}_{0}(m^{2}_{K})}\cdot\frac{m^{2}_{D}-m^{2}_{K}}{m^{2}_{D}-m^{2}_{\pi}}
  \; , \\ \\
& \displaystyle
1+\frac{a_{1}}{a_{2}}\cdot\frac{g_{\rho}}{f_{K}}\cdot\frac{F^{BK}_{1}(m^{2}_{\rho})}
{A^{B\rho}_{0}(m^{2}_{K})}\cdot\frac{(\varepsilon^{*}_{\rho}\cdot p^{~}_{D})}
{(\varepsilon^{*}_{\rho}\cdot p^{~}_{K})} \; , \\ \\
& \displaystyle
1+\frac{a_{1}}{a_{2}}\cdot\frac{g_{a_{1}}}{f_{K}}\cdot\frac{F^{BK}_{1}(m^{2}_{a_{1}})}
{A^{Ba_{1}}_{0}(m^{2}_{K})}\cdot\frac{(\varepsilon^{*}_{a_{1}}\cdot p^{~}_{D})}
{(\varepsilon^{*}_{a_{1}}\cdot p^{~}_{K})} \; ,
\end{array}
\right .
\eqno{(6)}
$$
where $V_{ij}$ ($i=u,c,t$; $j=d,s,b$) are the Cabibbo-Kobayashi-Maskawa (CKM)
matrix elements;
$\varepsilon$ denotes the polarization vector of $1^{\pm}$ mesons, $p$ stands
for the
4-momentum of a particle, and $a_{1,2}$ are the effective Wilson coefficients.
The decay constants and formfactors in eq. (6) are self-explanatory and their
values can be found from refs. [2,9]. From the above results we observe
$|T_{b}/T_{a}|\sim 1/\lambda^{2}\approx 20$ and
Im$(T_{b}/T_{a})\sim A^{2}\lambda^{2}\eta < 1/20$, where $\lambda, A$ and
$\eta$ are the Wolfenstein
parameters of the CKM matrix [10]. This implies that the Re$(\epsilon)$ term
of $\Delta_{K_{S}}$ plays the dominant role in ${\cal A}(X^{\pm}K_{S})$, even
though the
size of $\sin (\delta_{b}-\delta_{a})$ might be maximal ($\pm 1$). To a good
degree of accuracy, one finds
$$
{\cal A}(\pi^{\pm}K_{S}) \; \approx \; {\cal A}(\rho^{\pm}K_{S}) \; \approx \;
{\cal A}(a^{\pm}_{1}K_{S}) \; \approx \; 2|\epsilon|\cos\phi_{\epsilon} \; .
\eqno{(7)}
$$
The same result is obtainable for the $CP$ asymmetries in $D^{\pm}\rightarrow
(\pi^{\pm}, \rho^{\pm}, a^{\pm}_{1})+K_{L}$.

\vspace{0.2cm}

With the help of experimental data [2] on $D^{+}\rightarrow (\pi^{+}, \rho^{+},
a^{+}_{1})+\bar{K}^{0}$,
the branching ratios of $D^{\pm}\rightarrow (\pi^{\pm}, \rho^{\pm},
a^{\pm}_{1})+K_{S}$ ($K_{L}$)
are estimated to be $1.4\%$, $3.3\%$ and $4.0\%$, respectively. As discussed
for the case of
semileptonic $D^{\pm}$ decays, here one can also consider the possibility to
sum over the available final states
in order to obtain a statistically significant signal of $CP$ violation.
Optimistically speaking,
only about $10^{6-7}$ $D^{\pm}$ events are needed to establish ${\cal
A}(\pi^{\pm}K_{S})$ under a
perfect experimental environment.

\vspace{0.2cm}

{\large\it Example 3}. $~$ Let us briefly discuss the effect of
$K^{0}-\bar{K}^{0}$ mixing
on $CP$ asymmetries in multi-body nonleptonic decays of the type
$D^{\pm} \rightarrow \pi^{\pm} + n_{0}\pi^{0} + n_{\pm}(\pi^{\pm}\pi^{\mp}) +
(K_{S}, K_{L})$,
where the integers $n_{0}$ and $n_{\pm}$ are not simultaneously vanishing. Such
decay modes can occur
through the same quark diagrams as $D^{\pm}\rightarrow \pi^{\pm}+(K_{S},K_{L})$
do, but
$(u\bar{u})$ or $(d\bar{d})$ pair(s) need be created from the vacuum to form
the additional $\pi^{0}$ or $(\pi^{+}\pi^{-})$ meson(s) in the final state.
It is easy to show that the $CP$ asymmetries in this type of decays are given
by
${\cal A}(\pi^{\pm}K_{S})$ or ${\cal A}(\pi^{\pm}K_{L})$, i.e.,
2Re($\epsilon$). Current
experimental data have reconstructed the modes $D^{+}\rightarrow
\bar{K}^{0}\pi^{+}\pi^{0}$,
$\bar{K}^{0}\pi^{+}\pi^{+}\pi^{-}$ and
$\bar{K}^{0}\pi^{+}\pi^{+}\pi^{-}\pi^{0}$, whose
branching ratios are $9.7\%$, $7.0\%$ and $5.4\%$ respectively [2]. Thus there
exists large
potential to isolate a signal of $CP$ violation in these multi-body processes.
For this purpose,
the required number of $D^{\pm}$ events might be of order $10^{6-7}$.

\vspace{0.2cm}

The above results show that the effect of $K^{0}-\bar{K}^{0}$ mixing on $CP$
asymmetries in weak $D^{\pm}$
decays is not negligible {\it a priori}. In a similar way one may discuss
Re($\epsilon$)-induced
$CP$ violation in the decay modes of $D^{\pm}_{s}$ or $\stackrel{(-)}{D}$$^{0}$
mesons. A $\tau$-charm
factory running at the $\psi^{''}$(3770) resonance with an integrated
luminosity ${\cal L}=10^{33}$
cm$^{-2}$s$^{-1}$ could produce about $4\times 10^{7}$ $D^{\pm}$ pairs [11],
just the number of events
needed for probing $CP$ violation at the level of $10^{-3}$. This amount of
$D^{\pm}$ events could
also be accumulated at the forthcoming $B$ factories, $Z$ factory, or hadron
machines. Thus it is
very promising to observe Re($\epsilon$)-induced $CP$ violation in exclusive
decays of $D^{\pm}$ mesons
in the near future. To finally explore $CP$ violation in the $c$-quark sector
or to
pin down new physics at the level of $10^{-3}$ in the $D$-meson system, one has
to study those weak $D$ transitions
where $K^{0}-\bar{K}^{0}$ mixing is absent or its effect can be safely
neglected.

\vspace{0.2cm}

In the following we discuss Re($\epsilon$)-induced $CP$ violation in the
semi-inclusive
and exclusive decays $B_{d}\rightarrow X(c\bar{c})+(K_{S},K_{L})$, whose quark
diagrams are shown in fig. 2.
The tree-level amplitude is dominant over the loop-induced penguin (hairpin)
amplitude, since the latter
can only occur through three-gluon exchanges. With the help of unitarity of the
CKM matrix, we obtain the ratio of transition amplitudes for
$\stackrel{(-)}{B}$$^{0}_{d}\rightarrow
X(c\bar{c})+K_{S}$:
$$
\begin{array}{lll}
\xi & = & \displaystyle\frac{A (\bar{B}^{0}_{d}\rightarrow X(c\bar{c})+K_{S} )}
{A (B^{0}_{d}\rightarrow X(c\bar{c})+K_{S} )} \\ \\
& = & \displaystyle\frac{1-\epsilon^{*}}{1+\epsilon^{*}}\cdot
\frac{(V_{cb}V^{*}_{cs})A_{c}+(V_{tb}V^{*}_{ts})A_{t}}{(V^{*}_{cb}V_{cs})A_{c}+(V^{*}_{tb}V_{ts})A_{t}} \; ,
\end{array}
\eqno{(8a)}
$$
where $A_{c,t}$ are the hadronic amplitudes containing the strong phases.
Because $|A_{t}/A_{c}|$ is
expected to be rather smaller than unity, $\xi$ can be further written as
$$
\xi \; \approx \; 1 - 2\epsilon^{*} -2\chi
\eqno{(8b)}
$$
with $\chi\approx {\rm i} \lambda^{2}\eta (A_{t}/A_{c})$ to lowest-order
approximation.
The nonvanishing $\chi$ implies direct $CP$ violation in the decay amplitude.
We shall see later on that only ${\rm Re} (\chi)\approx -\lambda^{2}\eta {\rm
Im}(A_{t}/A_{c})$ enters the $CP$
asymmetries of our interest. For an exclusive decay mode like $B_{d}\rightarrow
J/\psi K_{S}$,
a rough estimate gives ${\rm Re}(\chi)\leq 10^{-3}$ [12]. It is likely that the
semi-inclusive decay rate of $B_{d}\rightarrow X(c\bar{c})+K_{S}$ should hardly
be affected by final-state
interactions (signified by $\chi$), since a sum over many available
$X(c\bar{c})$ states may give rise to
dilution of different $\chi$.

\vspace{0.2cm}

Specifically we assume that the experimental scenario is at the $\Upsilon (4S)$
resonance, the basis of the
forthcoming $B$-meson factories. On the $\Upsilon (4S)$ resonance, the $B$'s
are produced
in a two-body ($B^{+}_{u}B^{-}_{u}$ or $B^{0}_{d}\bar{B}^{0}_{d}$) state with
{\it odd} charge-conjugation
parity. Since the two neutral $B$ mesons mix coherently untill one of them
decays, one can
use the semileptonic decay of one meson to tag the flavor of the other meson
decaying to
$X(c\bar{c})K_{S}$ or $X(c\bar{c})K_{L}$. A generic formalism for the
time-dependent or time-integrated decays of
any coherent $P^{0}\bar{P}^{0}$ system ($P^{0}=K^{0}, D^{0}, B^{0}_{d}$ or
$B^{0}_{s}$) has been
given by the author in ref. [13]. For our present purpose we only consider the
time-integrated
$B^{0}_{d}\bar{B}^{0}_{d}$ transitions, which can be measured at either
symmetric or asymmetric $B$
factories. The joint decay rates of $(B^{0}_{d}\bar{B}^{0}_{d})_{\Upsilon
(4S)}\rightarrow
(X(c\bar{c})+K_{S})_{B_{d}}(l^{\pm}+ ...)_{B_{d}}$ are given as [13]
$$
\begin{array}{lll}
{\rm R}(l^{+}, X(c\bar{c})K_{S}) & \propto & \displaystyle |p/q|^{2} +
|\xi|^{2}
- a \left (|p/q|^{2} -|\xi|^{2}\right ) \; , \\
{\rm R}(l^{-}, X(c\bar{c})K_{S}) & \propto & \displaystyle |q/p|^{2}\cdot \left
[
|p/q|^{2} + |\xi|^{2} + a \left (|p/q|^{2} -|\xi|^{2}\right ) \right ] \; ,
\end{array}
\eqno{(9)}
$$
where $q/p\equiv (1-\epsilon^{~}_{B})/(1+\epsilon^{~}_{B})$ is a mixing
parameter of the
$B^{0}_{d}-\bar{B}^{0}_{d}$ system, and $a=(1-y^{2}_{d})/(1+x^{2}_{d})$ with
$x_{d}\equiv \Delta m^{~}_{B}/\Gamma_{B}$
and $y_{d}\equiv \Delta \Gamma_{B}/(2\Gamma_{B})$. The current experimental
data give
$x_{d} \approx 0.71$ [2], while a model-independent analysis shows $y_{d}\leq
10^{-2}$ [3]. In the context of
the standard model one estimates $2{\rm Re}(\epsilon^{~}_{B})\sim O(10^{-3})$
[6], which is of the same order
as $|\epsilon|$. Thus $y^{2}_{d}$ is negligible in the decay-rate difference
between
${\rm R}(l^{+}, X(c\bar{c})K_{S})$ and ${\rm R}(l^{-}, X(c\bar{c})K_{S})$. To
lowest-order approximations of $\epsilon^{~}_{B}$, $\epsilon$ and $\chi$, we
obtain the time-independent $CP$ asymmetry
$$
\begin{array}{lll}
{\cal A}_{CP} & = & \displaystyle\frac{{\rm R}(l^{-}, X(c\bar{c})K_{S}) - {\rm
R}(l^{+}, X(c\bar{c})K_{S})}
{{\rm R}(l^{-}, X(c\bar{c})K_{S}) + {\rm R}(l^{+}, X(c\bar{c})K_{S})} \\ \\
& \approx & \displaystyle\frac{2{\rm Re}(\epsilon) - 2x^{2}_{d}{\rm
Re}(\epsilon^{~}_{B}) +2{\rm Re}(\chi)}
{1+x^{2}_{d}} \; .
\end{array}
\eqno{(10)}
$$
As pointed out before, ${\rm Re}(\chi)$ might be comparable in magnitude with
${\rm Re}(\epsilon)$
and ${\rm Re}(\epsilon^{~}_{B})$ in the exclusive decay modes such as
$B_{d}\rightarrow J/\psi K_{S}$.
For the semi-inclusive decay $B_{d}\rightarrow X(c\bar{c})+K_{S}$, the
contribution of
${\rm Re}(\chi)$ should be less important due to possible dilution effects. If
this is true,
then the semi-inclusive asymmetry ${\cal A}_{CP}$ mainly measures $CP$
violation in the
$K^{0}-\bar{K}^{0}$ and $B^{0}_{d}-\bar{B}^{0}_{d}$ mixing matrices. Note that
the same $CP$
asymmetry (${\cal A}_{CP}$) is obtainable for $B_{d}\rightarrow
X(c\bar{c})+K_{L}$.

\vspace{0.2cm}

In principle ${\rm Re}(\epsilon^{~}_{B})$ can be measured from the charge
asymmetry of semileptonic
$B_{d}$ decays on the $\Upsilon (4S)$ resonance [6]: ${\cal A}_{\rm SL}\approx
4{\rm Re}(\epsilon^{~}_{B})$.
In practice to measure this signal to 3 standard deviations requires
$10^{7}$ like-sign dilepton events, corresponding to about $10^{9-10}$
$B^{0}_{d}\bar{B}^{0}_{d}$ events.
So many events are only reachable within the second-round experiments at a $B$
factory [3]. In
comparison, to measure the $CP$ asymmetry ${\cal A}_{CP}$ needs a bit more
$B^{0}_{d}\bar{B}^{0}_{d}$ events due to the cost for flavor tagging. It has
been shown
that the semi-inclusive branching ratio of $B_{d}\rightarrow X(c\bar{c})K_{S}$
is about
$1\%$ [14]. Practically one may use the semileptonic decays
$B^{0}_{d}\rightarrow
D^{*-}l^{+}\nu^{~}_{l}$ and $\bar{B}^{0}_{d}\rightarrow
D^{*+}l^{-}\bar{\nu}^{~}_{l}$, which have
branching ratios of $4.4\%$ [2], to tag the flavor of $B_{d}$ mesons decaying
to
$X(c\bar{c})+(K_{S},K_{L})$. Thus it is possible to establish ${\cal A}_{CP}$
to 3 standard deviations
with about $10^{10-11}$ $B^{0}_{d}\bar{B}^{0}_{d}$ events on the $\Upsilon
(4S)$ resonance.
Here the interesting point is that smallness of ${\cal A}_{CP}$ implies its
high
sensitivity to new physics in $\Delta B=2$ transitions [15] or from $CPT$
violation [16].
For this reason, a more detailed study of $CP$ violation in the semi-inclusive
and exclusive
decays $B_{d}\rightarrow X(c\bar{c})+(K_{S}, K_{L})$ is worthwhile.

\vspace{0.35cm}

The author would like to thank Professor H. Fritzsch for his warm hospitality
and constant encouragement.
He is also grateful to Professor A. I. Sanda for an enlightening comment on
this work, and to Professors
J. Bernab$\rm\acute{e}$u, D. Du, P. Wang, D. D. Wu and J. M. Wu for several
helpful conversations.
He finally acknowledges the Alexander von Humboldt Foundation for its financial
support.

\normalsize

\newpage

\begin{figure}
\begin{picture}(400,250)
\put(70,240){\line(1,0){90}}
\put(70,190){\line(1,0){90}}
\put(62,238){$c$}
\put(62,186){$\bar{d}$}
\put(43,210.5){$D^{+}$}
\put(163,238){$d$}
\put(163,186){$\bar{d}$}
\put(163,223){$\bar{s}$}
\put(163,200){$u$}
\put(180,193){$\pi^{+}$}
\put(180,230){$K^{0}$}
\put(160,215){\oval(70,25)[l]}
\put(85,240){\vector(1,0){2}}
\put(85,190){\vector(-1,0){2}}
\put(145,240){\vector(1,0){2}}
\put(145,190){\vector(-1,0){2}}
\put(145,227.5){\vector(-1,0){2}}
\put(145,202.5){\vector(1,0){2}}
\multiput(110,240)(3,-5){5}{\line(0,-1){5}}
\multiput(107,240)(3,-5){6}{\line(1,0){3}}
\put(113,165){(a1)}
\put(262,229){$c$}
\put(262,197.5){$\bar{d}$}
\put(243,210.5){$D^{+}$}
\put(363,236){$u$}
\put(363,189){$\bar{s}$}
\put(363,202.5){$d$}
\put(363,222){$\bar{d}$}
\put(380,228){$\pi^{+}$}
\put(380,194){$K^{0}$}
\put(270,215){\oval(60,31)[r]}
\put(360,215){\oval(52,21)[l]}
\put(360,215){\oval(77,46)[l]}
\put(285,230){\vector(1,0){2}}
\put(285,200){\vector(-1,0){2}}
\put(345,225){\vector(-1,0){2}}
\put(345,204.5){\vector(1,0){2}}
\put(345,238){\vector(1,0){2}}
\put(345,192){\vector(-1,0){2}}
\multiput(299.2,215)(5.3,0){4}{\small $\wedge$}
\put(313,165){(a2)}
\end{picture}

\begin{picture}(400,130)
\put(70,215){\line(1,0){90}}
\put(62,213){$c$}
\put(62,186){$\bar{d}$}
\put(42,200){$D^{+}$}
\put(163,212){$s$}
\put(163,186){$\bar{d}$}
\put(163,244){$u$}
\put(163,230){$\bar{d}$}
\put(180,236){$\pi^{+}$}
\put(180,199){$\bar{K}^{0}$}
\put(70,190){\line(1,0){90}}
\put(160,240){\oval(70,15)[l]}
\put(145,247.5){\vector(1,0){2}}
\put(145,232.5){\vector(-1,0){2}}
\put(85,215){\vector(1,0){2}}
\put(145,215){\vector(1,0){2}}
\put(85,190){\vector(-1,0){2}}
\put(145,190){\vector(-1,0){2}}
\multiput(110,215)(3,5){5}{\line(0,1){5}}
\multiput(107,215)(3,5){6}{\line(1,0){3}}
\put(113,165){(b1)}
\put(270,240){\line(1,0){90}}
\put(270,190){\line(1,0){90}}
\put(262,238){$c$}
\put(262,186){$\bar{d}$}
\put(243,210.5){$D^{+}$}
\put(363,238){$s$}
\put(363,186){$\bar{d}$}
\put(363,223){$\bar{d}$}
\put(363,200){$u$}
\put(380,193){$\pi^{+}$}
\put(380,230){$\bar{K}^{0}$}
\put(360,215){\oval(70,25)[l]}
\put(285,240){\vector(1,0){2}}
\put(285,190){\vector(-1,0){2}}
\put(345,240){\vector(1,0){2}}
\put(345,190){\vector(-1,0){2}}
\put(345,227.5){\vector(-1,0){2}}
\put(345,202.5){\vector(1,0){2}}
\multiput(310,240)(3,-5){5}{\line(0,-1){5}}
\multiput(307,240)(3,-5){6}{\line(1,0){3}}
\put(313,165){(b2)}
\end{picture}
\vspace{-4.8cm}
\caption{Quark diagrams for $D^{+}\rightarrow \pi^{+} K_{S}$ or $\pi^{+} K_{L}$
in the standard model.
Here $\pi^{+}$ can be replaced by $\rho^{+}$ or $a^{+}_{1}$.}
\end{figure}
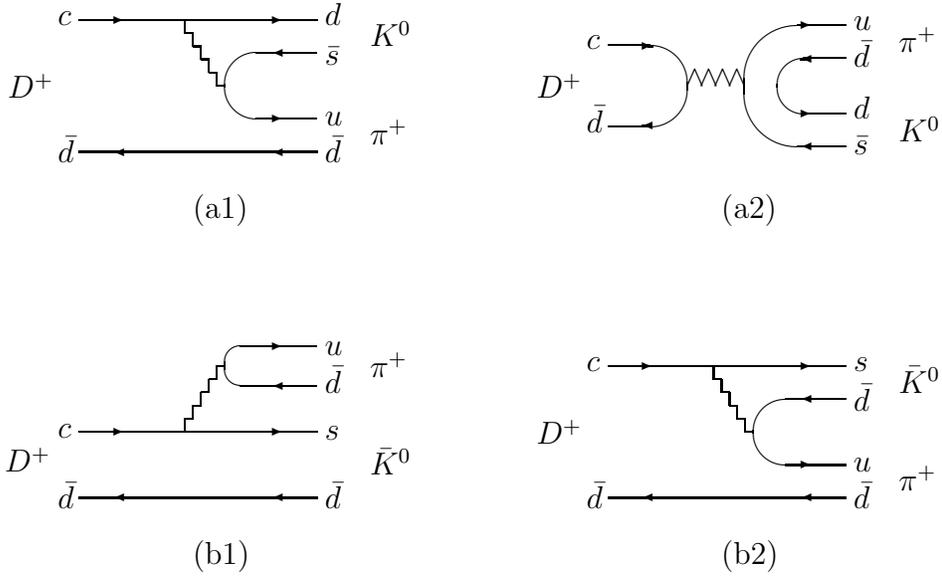

\begin{figure}
\begin{picture}(400,365)(20,0)
\put(70,330){\line(1,0){90}}
\put(70,280){\line(1,0){90}}
\put(160,305){\oval(90,25)[l]}
\put(145,330){\vector(-1,0){2}}
\put(163,325){$\bar{c}$}
\put(145,280){\vector(1,0){2}}
\put(163,275){$d$}
\put(85,330){\vector(-1,0){2}}
\put(62,325){$\bar{b}$}
\put(145,317.5){\vector(1,0){2}}
\put(163,312.5){$c$}
\put(85,280){\vector(1,0){2}}
\put(62,275){$d$}
\put(145,292.5){\vector(-1,0){2}}
\put(163,287.5){$\bar{s}$}
\put(42,300){$B^{0}_{d}$}
\put(172,320){$X(c\bar{c})$}
\put(172,282){$K^{0} \Longrightarrow K_{S,L}$}
\multiput(100,330)(3,-5){5}{\line(0,-1){5}}
\multiput(97,330)(3,-5){6}{\line(1,0){3}}
\put(113,255){(a)}
\put(310,305){\line(1,0){30}}
\put(370,305){\line(1,0){30}}
\put(355,305){\oval(30,25)[b]}
\put(310,280){\line(1,0){90}}
\put(400,335){\oval(90,25)[l]}
\put(325,305){\vector(-1,0){2}}
\put(302,302){$\bar{b}$}
\put(325,280){\vector(1,0){2}}
\put(302,275){$d$}
\put(282,290){$B^{0}_{d}$}
\put(385,305){\vector(-1,0){2}}
\put(403,302){$\bar{s}$}
\put(385,280){\vector(1,0){2}}
\put(403,275){$d$}
\put(385,347.5){\vector(-1,0){2}}
\put(403,345.5){$\bar{c}$}
\put(385,322.5){\vector(1,0){2}}
\put(403,320.5){$c$}
\put(412,333){$X(c\bar{c})$}
\put(412,289){$K^{0} \Longrightarrow K_{S,L}$}
\multiput(339,305)(6,0){5}{$\wedge$}
\put(353,255){(b)}
\end{picture}
\vspace{-7.9cm}
\caption{Tree-level and penguin (hairpin) diagrams for the semi-inclusive (or
exclusive)
decay modes $B^{0}_{d}\rightarrow X(c\bar{c}) + (K_{S}, K_{L})$.}
\end{figure}
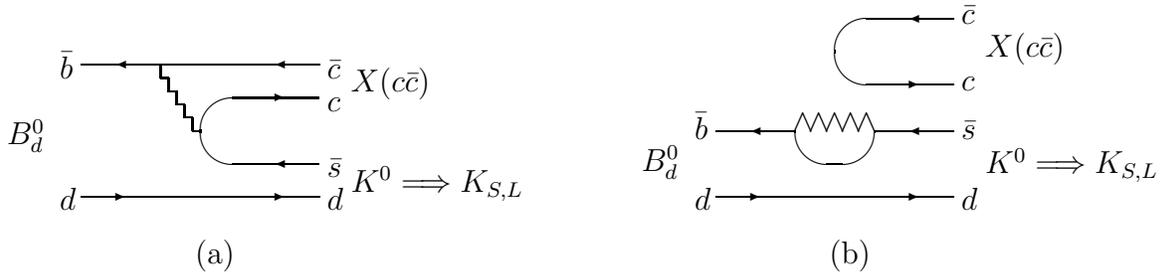


\vspace{0.6cm}

\begin{thebibliography}{99}

\bibitem{1} J. H. Christenson, et al., Phys. Rev. Lett. 13 (1964) 138.

\bibitem{2} Particle Data Group, L. Montanet et al., Phys. Rev. D50 (1994)
1173.

\bibitem{3} For recent reviews and extensive references, see the Proceedings of
1994 International
Workshop on $B$ Physics and Physics Beyond the Standard Model at the $B$
Factory, Nagoya, October 1994
(edited by A. I. Sanda).

\bibitem{4} See, e.g., A. Pich, preprint CERN-TH.7066/93 (1993); \\
J. R. Fry and T. Ruf, preprint CERN-PPE/94-20 (1994); \\
J. L. Hewett, preprint SLAC-PUB-6695 (1994); \\
Tiehui Liu, preprint HUTP-94/E021 (1994); and references therein.

\bibitem{5} For recent reviews and extensive references, see the proceedings of
{\it Beauty' 93}
and {\it Beauty' 94} workshops (Nucl. Inst. and Meth. A351 (1994) $\&$ A333
(1993)).

\bibitem{6} I. I. Bigi, V. A. Khoze, N. G. Uraltsev, and A. I. Sanda,
in {\it CP Violation}, edited by C. Jarlskog (World Scientific, Singapore,
1988), p. 175; \\
I. I. Bigi, in the Proceedings of the Tau-Charm Factory Workshop, Stanford, CA,
May 1989 (edited by
L. V. Beers), p. 169.

\bibitem{7} F. Buccella et al., Phys. Lett. B302 (1993) 319; preprint
Hep-Ph/9411286 (1994).

\bibitem{8} I. I. Bigi and H. Yamamoto, Phys. Lett. B349 (1995) 363; \\
M. Golden and B. Grinstein, Phys. Lett. B222 (1989) 501.

\bibitem{9} A. J. Buras, J. M. G$\rm\acute{e}$rard, and R. R$\rm\ddot{u}$ckl,
Nucl. Phys. B268 (1986) 16; \\
M. Bauer, B. Stech, and M. Wirbel, Z. Phys. C23 (1987) 103; \\
M. Neubert et al., in {\it Heavy Flavours}, edited by A. J. Buras and M.
Lindner (World Scientific, Singapore, 1992), p. 286.

\bibitem{10} L. Wolfenstein, Phys. Rev. Lett. 51 (1983) 1945; \\
M. Kobayashi, Prog. Theor. Phys. 92 (1994) 287; \\
Z. Z. Xing, Phys. Rev. D51 (1995) 3958.

\bibitem{11} J. Bernab$\rm\acute{e}$u, preprint TRI-PP-93-100 $\&$ FTUV/93-22
(1993).

\bibitem{12} D. Du and Z. Z. Xing, Phys. Lett. B312 (1993) 199.

\bibitem{13} Z. Z. Xing, preprint LMU-22/94 (1994).

\bibitem{14} J. Bernab$\rm\acute{e}$u and C. Jarlskog, Phys. Lett. B301 (1993)
275.

\bibitem{15} For reviews and extensive references, see: \\
{\it CP Violation}, edited by C. Jarlskog (World Scientific, Singapore, 1988);
\\
{\it CP Violation}, edited by L. Wolfenstein (North-Holland, 1989).

\bibitem{16} M. Kobayashi and A. I. Sanda, Phys. Rev. Lett. 69 (1992) 3139; \\
Z. Z. Xing, Phys. Rev. D50 (1994) R2957; \\
D. Colladay and V. A. Kosteleck$\rm\acute{y}$, Phys. Lett. B344 (1995) 259; \\
D. D. Wu, preprint PVAMU-HEP-11-94 (1994).

\end{thebibliography}
\end{document}